\documentclass{llncs}

\usepackage{makeidx}  
\usepackage{graphicx}
\newcommand{\bra}[1]{\left(#1\right)}

\begin{document}

\frontmatter          

\pagestyle{headings}  

\title{Statistical Neurodynamics for sequence processing neural networks with finite dilution\thanks{Lect. Notes Comput. Sci. 4491, Part I, pp. 1148¨C1156, 2007.(ISNN 2007)}}

\author{Pan Zhang \thanks{Email: {\tt july.lzu@gmail.com}} \and Yong Chen\thanks{Corresponding author. Email: {\tt ychen@lzu.edu.cn}}}

\institute{Institute of Theoretical Physics, Lanzhou University, 730000 Lanzhou, China\\
}

\maketitle              

\begin{abstract}
We extend the statistical neurodynamics to study transient dynamics of sequence processing neural networks with finite dilution, and the theoretical results are supported by extensive numerical simulations. It is found that the order parameter equations are completely equivalent to those of the Generating Functional Method, which means that crosstalk noise follows normal distribution even in the case of failure in retrieval process. In order to verify the gaussian assumption of crosstalk noise, we numerically obtain the cumulants of crosstalk noise, and third- and fourth-order cumulants are found to be indeed zero even in non-retrieval case.
\end{abstract}

\section{Introduction}

Models of attractor neural networks for processing sequences of
patterns, as a realization of a temporal association, have been of
great interest over some time
\cite{seq_1,During98,Kawamura02,Theumann03,yong00}. On contrary to
Hopfield model \cite{Hopfield,AGS}, the asymmetry of the interaction
matrix in this model leads to violation of detailed balance, ruling
out an equilibrium statistical mechanics analysis. Usually,
Generating Functional Method and Statistical Neurodynamics are two
ways to study this model. Generating Functional Method
\cite{During98,gfa1,gfa2,gfa3} is also known as e.g. path integral
method, and it allows the exact solution of the dynamics and to
search all relevant physical order parameters at any time step via
the derivatives of a generating functional. Statistical
Neurodynamics (so called Signal-To-Noise Analysis)
\cite{Kawamura02,Amari,Masato1,spars-seq}, starts from splitting
of the local field into a signal part originating from the pattern
to be retrieved and a noise part arising from the other patterns,
and uses simple assumption that crosstalk noise is normally
distributed with zero mean and evaluated variance. In Hopfield
model, this treatment is proved to be just an approximation and the
Gaussian form of crosstalk noise only holds when pattern recall
occurs \cite{Nishimori-Ozeki,Nishimori2,coolen1}. But Kawamuran et
al. suggest that Statistical Neurodynamics is exact in fully
connected sequence processing model \cite{Kawamura02}. In this paper
we show that treatment of Statistical Neurodynamics is also exact
correct in the case of diluted connection.

In the nature of brain, fully-connected structure of network is
biological unrealistic. Therefore, as a relaxation of conventional,
unrealistic condition that every neuron is connected to every other,
diluted networks have received lots of attention
\cite{Watkin91,Derrida87,Patrick90,Castillo,Sompolinsky1986}. There
are several kinds of dilution, with different average connections
$cN$, where $c$ is connection probability. If $cN=\textsl{O}\bra{\ln
n}$, it is called extreme dilution, and in this case, network has a
local Caley-Tree structure and almost all pairs of neurons have
entirely different sets of ancestors. Thus, the correlations in
noise may be neglected. The reported first that the extremely
diluted asymmetric network can be calculated exactly is Derrida et
al \cite{Derrida87}. Extremely diluted symmetric Hopfield network
was also studied using replica-symmetric calculation. When
$cN=\textsl{O}\bra{1}$, network only have finite connection
regardless of its scale, and the equilibrium properties has been
calculated using replica-symmetric approximation
\cite{finite_connectivity}. In this work, we focus on the case that
$cN=\textsl{O}(N)$, and we will show that our theoretical result
also suitable for extremely diluted network $cN\ll N$. Recently,
Theumann \cite{Theumann03} studied this type of dilution using
Generating Functional Method, and we will discuss the relationship
between our result and those obtained by Theumann.

This paper was organized as follows. In Section \ref{sec2} we recall
the sequence processing neural networks with finite dilution. In
Section \ref{sec3} we use statistical neurodynamics to obtain the
order parameter equations which describe dynamics of network, and
compare our theoretical result with numerical simulations. Section
\ref{sec4} discuss the relationship between our results and those
obtained by Generating Functional Method. Section \ref{sec5} contains
conclusion and discussion.

\section{Model definition \label{sec2}}

Let us consider a sequence processing model that consists of $N$ spins of neurons, when $N\rightarrow \infty$. The state of spins takes $S_i\bra{t}=\pm 1$ and updates the state synchronously with following probability \cite{chen01}:
\begin{equation}
\mathbf{Prob}[S_i\bra{t+1}|h_i\bra{t}]=\frac{1}{2}[1+S_i\bra{t+1}\tanh \beta
h_i\bra{t}],
\label{eq-1}
\end{equation}
where $\beta=1/T$ is inverse temperature and local field is given by
\begin{equation}
h_i\bra{t}=\sum_{j=1}^NJ_{ij}S_j\bra{t}.
\label{eq-2}
\end{equation}
When we use $F\bra{\cdot}$ to express transfer function, the parallel dynamics is expressed by
\begin{equation}
S_i\bra{t+1}=F(h_i\bra{t}).
\label{eq-3}
\end{equation}

We store $P=\alpha N$ random patterns $\xi^\mu=\bra{\xi_1^\mu,...,\xi_N^\mu}$ in network, where $\alpha$ is the loading ratio. Interaction matrix $J_{ij}$ is chosen to retrieve the patterns as $\xi^1\rightarrow\xi^2\rightarrow...\xi^P\rightarrow\xi^1$ sequentially. For instance, it is given by \cite{yong00}
\begin{equation}
J_{ij}=\frac{c_{ij}}{cN}\sum_{\mu=1}^P\xi_i^{\mu+1}\xi_j^\mu,
\label{eq-4}
\end{equation}
with $\xi^{\bra{P+1}}=\xi^1$. $c_{ij}$ takes its value as $0$ or $1$
probabilities
\begin{eqnarray}
\mathbf{Prob}\bra{c_{ij}=1}=1-\mathbf{Prob}\bra{c_{ij}=0}=c
\label{eq-5}
\end{eqnarray}
with $\bra{0\leq c\leq 1}$ and $c_{ij}=c_{ji}$ for symmetric dilution.

Let us consider the case to retrieve $\xi^q$. We define $m^q\bra{t}$
as the overlap parameter between the network state
$\mathbf{S}\bra{t}$ and the condensed pattern $\mathbf{\xi}^q$
\begin{equation}
m^q\bra{t}=\frac{1}{N}\sum_{i=1}^N\xi_i^qS_i\bra{t}. \label{eq-6}
\end{equation}

\section{Statistical Neurodynamic for diluted networks \label{sec3}}

We start from Somplinsky's idea that in the limit
$N\rightarrow\infty$ the synaptic matrix in (\ref{eq-4}) can be
written as a fully connected model with synaptic
noise \cite{Sompolinsky1986}
\begin{equation}\label{som}
    J_{ij}^{\mathtt{eff}}=\frac{1}{N}\sum_{\mu=0}^{\alpha N}\xi_i^{\mu+1}
    \xi_j^\mu+\eta_{ij},
\end{equation}
where $\eta_{ij}$ is a complex random variable following Gaussian
distribution with mean $0$ and variance $\alpha (1-c)/c$. Theumann
also proved this relationship by taking expansion of averaging
$e^{-i\mathbf{\hat{h}}\bra{t}\cdot \mathbf{J}\mathbf{S}\bra{t}}$
over disorder, where $\mathbf{\hat{h}}$ is auxiliary local field
(see section \ref{sec4}).

Then local field is decribed by
\begin{equation}\label{h1}
    h_i\bra{t}=\xi_i^{q+1}m^q\bra{t}+Z_i\bra{t}.
\end{equation}
Here $Z_i\bra{t}$ denotes the crosstalk noise term from uncondensed patterns,
\begin{equation}\label{z1}
    Z_i\bra{t}=\frac{1}{N}\sum_{\mu\neq q}^{\alpha N}\sum_{j\neq
    i}^N\xi_i^{\mu+1}\xi_j^\mu
    S_j\bra{t}+\sum_{j=1}^NS_j\bra{t}\eta_{ij}.
\end{equation}

We assume that the noise term $Z_i\bra{t}$ is normally distributed with mean $0$ and variance $\sigma_i^2\bra{Z_i\bra{t}}$. For Hopfield model, the assumption is shown to be valid within statistical errors by Monte Carlo simulations as long as the memory retrieval is successful \cite{Nishimori-Ozeki}. And for sequence processing neural networks, the assumptions is shown to be hold even in the non-retrieval case \cite{Kawamura02}. We also keep this assumption in our model with presence of $c_{ij}$, which is verified by numerical simulations of cumulants of $Z_i\bra{t}$.

To obtain the variance $\sigma^2\bra{Z_i\bra{t+1}}$ in a recursive form, we express the noise term of the transfer function $F\bra{\cdot}$,
\begin{eqnarray}\label{noise1}
    Z_i\bra{t+1}&=&\frac{1}{N}\sum_{\mu\neq q}^{\alpha N}\sum_{j\neq i}^N\xi_i^{\mu+1}\xi_j^\mu
    F\bra{h_j\bra{t}}+\sum_{j=1}^NF\bra{h_j\bra{t}}\eta_{ij} \nonumber\\
    &=&Y_i\bra{t+1}+\sum_{j=1}^NF\bra{h_j\bra{t}}\eta_{ij},
\end{eqnarray}
where mean of $Y_i\bra{t+1}$ is zero, and
\begin{equation}\label{h2}
    h_j\bra{t}=\xi_j^{q}m^{q-1}\bra{t}+\frac{1}{N}\sum_{k\neq j}^N \sum_{\nu\neq q-1}^{\alpha N} \xi_j^{\nu+1}\xi_k^\nu S_k\bra{t-1} + \sum_j^N S_k\bra{t-1}\eta_{jk}.
\end{equation}
Since $h_j\bra{t}$ depends strongly on $\xi_j^\mu$, and the term
$\frac{1}{N}\sum_{k\neq j}^N\xi_j^\mu\xi_k^{\mu-1}S_k\bra{t}$ can be
rationally considered as stochastic order $\textsl{O}(1/\sqrt{N})$,
one can expand the transfer function $F\bra{\cdot}$ up to the first
order, then obtain
\begin{equation}\label{Y1}
    Y_i\bra{t+1}=\frac{1}{N}\sum_{j\neq i}^N\sum_{\mu\neq q}^{\alpha
    N}\xi_i^{\mu+1}\xi_j^\mu
    F\bra{\hat{h}_j\bra{t}}+U\bra{t+1}Y_i\bra{t},
\end{equation}
where
\begin{equation}\label{h3}
    \hat{h}_j\bra{t}=\xi_j^{q}m^{q-1}\bra{t}+\frac{1}{N}\sum_{k\neq
    j}^N\sum_{\nu\neq q-1,\mu}^{\alpha N}\xi_j^{\nu+1}\xi_k^\nu
    S_k\bra{t-1}+\sum_j^NS_k\bra{t-1}\eta_{jk}
\end{equation}
and
\begin{equation}\label{U(t)}
    U\bra{t+1}=\frac{1}{N}\sum_{j\neq i}^NF'\bra{\hat{h}_j\bra{t}}.
\end{equation}

Since crosstalk noise is assumed to be zero mean, one can calculate variance of crosstalk noise directly by
\begin{equation}\label{eq_variance_Z}
    \sigma^2\bra{Z_i\bra{t+1}}=\sigma^2\bra{Y_i\bra{t+1}}+\alpha
    (1-c)/c.
\end{equation}
Firstly, we have to determine $\sigma^2\bra{Y_i\bra{t+1}}$,
\begin{eqnarray}\label{eq_variance}
    \sigma^2\bra{Y_i\bra{t+1}}&=&\frac{1}{N^2}E\left[\sum_j\sum_\mu\bra{\xi_i^{\mu +1}}^2\bra{\xi_j^\mu}^2\bra{F\bra{\hat{h}_j\bra{t}}}^2\right]+U^2\bra{t+1} \sigma^2\bra{Y_i\bra{t}}\nonumber\\ &+&E\left[2\frac{U\bra{t+1}}{N} \sum_{j,k}\sum_{\mu,\nu}\xi_i^{\mu+1}\xi_i^{\nu+1}\xi_j^\mu
    \hat{S}_j\bra{t+1}\xi_k^{\nu-1}S_k\bra{t}\right],
\end{eqnarray}
where we use $\hat{S}_j\bra{t+1}=F\bra{\hat{h}_j\bra{t}}$. According
to the literature \cite{Kawamura02,spars-seq}, when expand
$S_k\bra{t}$ up to $t=0$, the last term in Eq. (\ref{eq_variance})
becomes
\begin{equation}\label{eq_vanishes}
    \frac{2}{N}E\left[ \sum_{n=1}^{t+1}\bra{\prod_{\tau=1}^nU\bra{t+1-\tau} \sum_j\hat{S}_j\bra{t+1}\hat{S}_j\bra{t+1-n}} \sum_\mu\bra{\xi_i^{\mu+1}}^2\xi_j^\mu\xi_j^{\mu-n} \right].
\end{equation}
Here all terms in Eq. (\ref{eq_vanishes}) are independent of
each other except for $n=\alpha N,2\alpha N,3\alpha N...$. So when
$N\rightarrow\infty$, the dependence vanishes, and the last term in
Eq. (\ref{eq_variance}) is ignored. The variance of $Y_i\bra{t+1}$
term becomes
\begin{equation}\label{eq_final_variance}
 \sigma_i^2\bra{Y_i\bra{t+1}}=\alpha +U^2\bra{t+1}\sigma_i^2\bra{Y_i\bra{t}}.
\end{equation}

When variance of crosstalk noise is determined using Eqs. (\ref{eq_variance_Z}, \ref{eq_final_variance}), all order parameters
can be expressed by the following closed equations
\begin{equation}\label{eq_m}
    m\bra{t+1}=\int Dz\left\langle \xi F\bra{\xi
    m\bra{t}+\sigma\bra{Z_i\bra{t}}z}\right\rangle_\xi,
\end{equation}
\begin{equation}\label{eq_U}
    U\bra{t+1}=\frac{1}{\sigma\bra{Z_i\bra{t}}}\int Dzz\left\langle F\bra{\xi
    m\bra{t}+\sigma\bra{Z_i\bra{t}}z}\right\rangle_\xi,
\end{equation}
\begin{equation}\label{eq_Z_final}
    \sigma^2\bra{Z_i\bra{t}}=\frac{\alpha}{c}+U^2\bra{t}\bra{\sigma^2\bra{Z_i\bra{t-1}}-\alpha \bra{1-c}/c},
\end{equation}
where $\langle\cdot\rangle_\xi$ stands for the average over the retrieval pattern $\xi$, and $Dz=\frac{1}{\sqrt{2\pi}}exp\bra{-z^2/2}dz$.

In the case that the temperature is absolute zero,
$F\bra{\cdot}=sgn\bra{\cdot}$, we get following expressions of order
parameters
\begin{equation}\label{eq_m_sgn}
    m\bra{t+1}=\mathtt{erf}\bra{\frac{m\bra{t}}{\sigma\bra{Z_i\bra{t}}}},
\end{equation}
\begin{equation}\label{eq_U_sgn}
    U\bra{t+1}=\frac{1}{\sigma\bra{Z_i\bra{t}}}\sqrt{\frac{2}{\pi}} \exp\bra{-\frac{m^2}{2\sigma^2\bra{Z_i\bra{t}}}},
\end{equation}
where
\begin{equation}\label{eq_erf}
    \mathtt {erf}\bra{u}=\sqrt{2/\pi}\int_0^u \exp\bra{-x^2/2}dx.
\end{equation}

\begin{figure}
\includegraphics[width=0.8\textwidth]{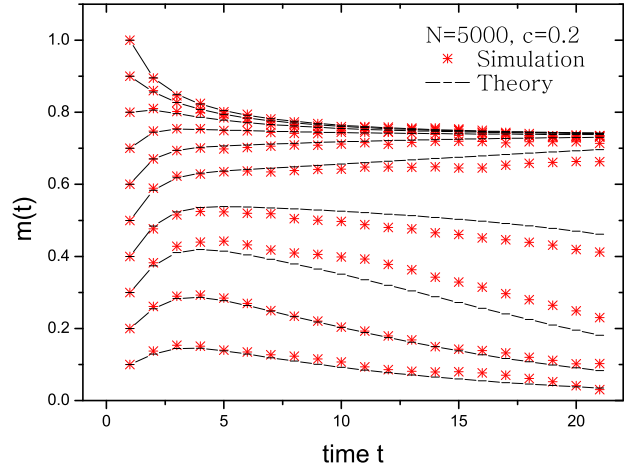}
\caption{\label{g_dynamics} Temporal evolution of overlap $m(t)$,
initial overlap ranging from $1.0$ to $0.1$ (up to down). The
parameters of networks are $N=5000$, $c=0.2$ and $\alpha/c=0.38$.}
\label{fig1}
\end{figure}

\begin{figure}
\includegraphics[width=0.8\textwidth]{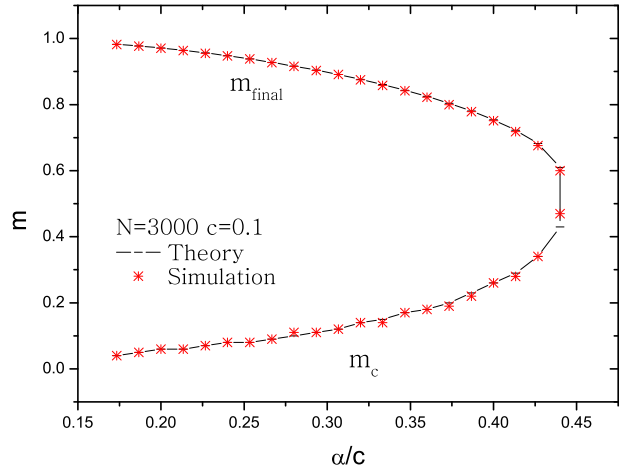}
\caption{\label{g_basin_of_attraction} Basin of attraction obtained by theory and numerical simulations. The parameters of networks are $N=5000$ and $c=0.1$.}
\label{fig2}
\end{figure}

This finishes statistical neurodynamics treatment of Sequence
Processing Neural Networks with finite synaptic dilution. The above
equations form a recursive scheme to calculate the dynamical
properties of the systems with an arbitrary time step. The time
evolution of overlaps obtained both in theory and numerical
simulation are plotted in Fig. \ref{fig1}. Initial overlaps range
from $0.1$ to $1.0$. when initial overlap is smaller than $0.5$,
network will fail to retrieval, and overlap will finally vanishes.
Fig. \ref{fig2} shows the basin of attraction both in theory and in
numerical simulations.

\section{Generating Functional Method and Statistical Neurodynamics \label{sec4}}

The idea of generating functional method is to concentrate on the moment generating function $Z\left[\psi\right]$, which fully captures the statistics of paths,
\begin{equation}\label{generating_functional}
    Z\left[\psi\right]=\sum_{\mathbf\sigma\bra{0},...,\mathbf\sigma\bra{t}}P\left[
    \mathbf\sigma\bra{0},...,\mathbf\sigma\bra{t}\right] e^{-i\sum_{s<t}\mathbf{S}\bra{s}\cdot\mathbf{\psi}\bra{s}}.
\end{equation}
The generating function $Z\left[\psi\right]$ involves the overlap parameter $m\bra{t}$. The response functions $G\bra{t,t'}$ and the correlation functions $C\bra{t,t'}$ are
\begin{equation}\label{eq_GFA_m}
    m\bra{t}=i\lim_{\mathbf{\psi}\rightarrow
    0}\frac{1}{N}\sum_{i=1}^{N}\xi_i^t\frac{\partial Z\left[\mathbf\psi\right]}{\partial
    \mathbf\psi_i\bra{t}}=\frac{1}{N}\sum_{i=1}^N\xi_i^t\langle
    S_i\bra{t}\rangle,
\end{equation}
\begin{equation}\label{eq_GFA_g}
    G\bra{t,t'}=i\lim_{\mathbf{\psi}\rightarrow
    0}\frac{1}{N}\sum_{i=1}^{N}\frac{\partial ^2 Z\left[\mathbf\psi\right]}{\partial
    \psi_i\bra{t}\partial\theta_i\bra{t'}}=\frac{1}{N}\sum_{i=1}^N\frac{\partial\langle
    S_i\bra{t}\rangle}{\partial\theta_i\bra{t'}},
\end{equation}
\begin{equation}\label{eq_GFA_c}
    C\bra{t,t'}=-\lim_{\mathbf{\psi}\rightarrow
    0}\frac{1}{N}\sum_{i=1}^{N}\frac{\partial ^2 Z\left[\mathbf\psi\right]}{\partial
    \psi_i\bra{t}\partial\psi_i\bra{t'}}=\frac{1}{N}\sum_{i=1}^N\langle
    S_i\bra{t}S_i\bra{t'}\rangle.
\end{equation}

D\"{u}ring et al. first discussed the sequence processing model using Generating Functional Method and obtained dynamical equations in the form of a multiple Gaussian integral \cite{During98}, which is too complex to calculate. Then Kawamura et al. simplified those equations and obtained a tractable description of dynamical equations with single Gaussian integral \cite{Kawamura02}. Theumann discussed the case of finite dilution using Generating Functional Method \cite{Theumann03}, with idea that
\begin{equation}\label{eq_average_e}
    \langle
e^{-i\hat{\mathbf{h}}\bra{s}\cdot\mathbf{J^{eff}}\mathbf{S}\bra{t}}\rangle_{c_{ij}}=
e^{-i\hat{\mathbf{h}}\bra{t}\cdot\mathbf{J}\mathbf{S}\bra{t}} e^{-\Delta^2\hat{\mathbf{h}}^2\bra{t}/2},
\end{equation}
where $\Delta^2=\alpha\bra{1-c}/c$ presents variance of independent
Gaussian random variables, which is exactly the idea of Sompolinsky
that $J_{ij}^{eff} = J_{ij} + \eta_{ij}$ \cite{Sompolinsky1986}.
With this formula, Theumann obtained the temporal evolution of order
parameters using the same scheme to that in \cite{During98,Kawamura02},
\begin{equation}\label{gfa_m}
    m\bra{t}=\int Dz\left \langle tanh\beta\left[m\bra{t-1}+\theta\bra{t-1}+z\sqrt{\alpha
    D\bra{t-1,t-1}}\right]\right\rangle_\xi,
\end{equation}
\begin{equation}\label{gfa_g}
    G\bra{t,t-1}=\beta\left\{1-\int Dz\left\langle tanh^2\beta\left[m\bra{t-1}+\theta\bra{t-1}+z\sqrt{\alpha
    D\bra{t-1,t-1}}\right]\right\rangle_\xi\right\},
\end{equation}
and $C\bra{t,t}=1$, where $Dz=1/\sqrt{2\pi}\exp\bra{-z^2/2}$. Covariance matrix of crosstalk noise is given by
\begin{equation}\label{gfa_d}
    D\bra{t,t}=R\bra{t,t}+\bra{1-c}/c,
\end{equation}
\begin{equation}\label{gfa_r}
    R\bra{t,t}=1+G^2\bra{t,t-1}R\bra{t-1,t-1}.
\end{equation}

\begin{figure}
\includegraphics[width=0.8\textwidth]{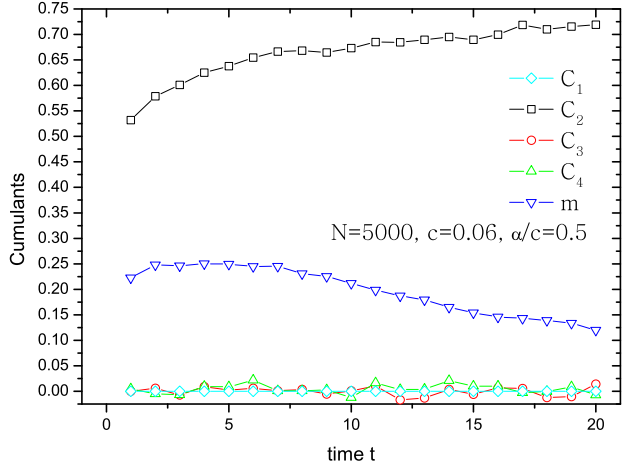}
\caption{\label{g_cumulants} Temporal evolution of cumulants $C_1\bra{t},C_2\bra{t},C_3\bra{t},C_4\bra{t}$, and overlap $m\bra{t}$. The initial overlap is $m\bra{0}=0.2$. The parameters of networks are $N=5000$, $c=0.06$, and $\alpha /c=0.5$.
\label{fig3}}
\end{figure}

From Eq. (\ref{eq_m}) to Eq. (\ref{gfa_m}), Note that
$\sigma^2\bra{Z_i\bra{t}}$ corresponds to $\alpha D\bra{t,t'}$, and
$U\bra{t}$ corresponds to $G\bra{t,t-1}$. It is easy to find that
Statistical Neurodynamics and Generating Functional Method present
the same order parameter equations of temporal evolution. It means
that the Gaussian form of crosstalk noise holds and Statistical
Neurodynamics can give the exact solution, comparing with Hopfield
model that the crosstalk noise is normally distributed only in
retrieval case \cite{Nishimori-Ozeki}. To verify the distribution of
crosstalk noise, the first, second, third, and fourth cumulants
$c_1\bra{t},c_2\bra{t},c_3\bra{t},c_4\bra{t}$ are evaluated
numerically, and the third and fourth cumulants are found to be zero
even when network fails in retrieval (see Fig. \ref{fig3}).

\section{Conclusion and discussion \label{sec5}}

In this paper, Statistical Neurodynamics is extended to study the retrieval dynamics of sequence processing neural networks with random synaptic dilution. Our theoretical results are exactly consistent with numerical simulations. The order parameter equations are obtained which is complete equivalent to those obtained by Generating Functional Method. We also present the first, second, third and forth-order cumulants of crosstalk noise to verify the Gaussian distribution of noise.

Finally, note that, in fully connected network or extremely diluted network, one can also obtain the order parameter equations from our Eqs. (\ref{eq_m}-\ref{eq_Z_final}). For $c=1$, $\sigma^2\bra{Z_i\bra{t}}=\sigma^2\bra{Y_i\bra{t}}$ and equation (\ref{eq_Z_final}) corresponds to the order parameter equations in fully connected networks \cite{Kawamura02}. For the limit $cN\ll N$, $\alpha \ll \alpha/c$, then the variance of crosstalk noise is always $\alpha/c$, that is exactly the result obtained in extremely diluted network where the local Cayley-Tree structure is held and all correlations in noise are neglected \cite{Derrida87}.

\section*{Acknowledgment}
The work reported in this paper was supported in part by the National Natural Science Foundation of China with Grant No. $10305005$ and the Special Fund for Doctor Programs at Lanzhou University.

\end{document}